\documentclass{article}
\pdfoutput=1

\usepackage[nonatbib,preprint]{nips_2018}

\usepackage[utf8]{inputenc} 
\usepackage[T1]{fontenc}    
\usepackage{hyperref}       
\usepackage{url}            
\usepackage{booktabs}       
\usepackage{amsfonts}       
\usepackage{nicefrac}       
\usepackage{microtype}      

\usepackage{graphicx}

\usepackage{amsmath}
\usepackage{amsfonts}
\usepackage{amssymb}
\newcommand{\myrightleftarrows}{\mathrel{\substack{\longrightarrow \\[-.6ex] \longleftarrow}}}
\usepackage{nicefrac}
\newtheorem{statement}{Statement}

\usepackage[inline]{enumitem}

\usepackage{tikz}
\usetikzlibrary{arrows,shapes,arrows, arrows.meta, positioning,shapes.geometric}

\usepackage[disable]{todonotes}

\newcommand{\wvec}{\mathbf{w}}

\newcommand{\md}[1]{\todo[color=green!40]{{#1}}}
\newcommand{\ma}[1]{\todo[color=pink!40]{{#1}}}
\usepackage{datetime}

\makeatletter
\def\bstctlcite{\@ifnextchar[{\@bstctlcite}{\@bstctlcite[@auxout]}}
\def\@bstctlcite[#1]#2{\@bsphack
  \@for\@citeb:=#2\do{%
    \edef\@citeb{\expandafter\@firstofone\@citeb}%
    \if@filesw\immediate\write\csname #1\endcsname{\string\citation{\@citeb}}\fi}%
  \@esphack}
\makeatother

\title{Pushing the boundaries of parallel Deep Learning -- A practical approach}

\author{
  Paolo Viviani \\
  Computer Science Dept.\\
  University of Torino, Italy\\
  Noesis Solutions NV \\
  \texttt{pviviani@unito.it} \\
\And
  Maurizio Drocco \\
  Pacific Northwest National Lab.\\
  \texttt{maurizio.drocco@pnnl.gov} \\
\And
  Marco Aldinucci \\
  Computer Science Dept.\\
  University of Torino, Italy\\
  \texttt{aldinuc@di.unito.it} \\
}

\begin{document}
\bstctlcite{IEEEexample:BSTcontrol}

\maketitle
\todo{NIPS requires 8 page excluding references. Acknowledgements are removed for blind review.}

\begin{abstract}
This work aims to assess the state of the art of data parallel deep neural 
network training, trying to identify potential research tracks to be exploited 
for performance improvement. Beside, it presents a design for a practical C++ 
library dedicated at implementing and unifying the current state of the art 
methodologies for parallel training in a performance-conscious framework, 
allowing the user to explore novel strategies without departing significantly 
from its usual work-flow.
\end{abstract}

\ma{Troppe volte dici literature, sostiuisci con i nomi degli autori a
  meno ch enon siano troppi paper diversi -- FATTO}
\section{Introduction}\label{sec:intro}
As deep learning techniques become more and more popular, there is the need to 
move these applications from the data scientist's Jupyter notebook to reliable 
and efficient enterprise solutions. This aim involves several steps to be taken, 
and this work advocates the need to push the current state of the art in 
parallel training in order to achieve:
\md{maybe better to inline the following list, so that the list below (i.e., research contributions) is more outstanding -- DONE}
\begin{enumerate*}
\item faster end-to-end training for large production datasets;
\item distributed training on the edge, namely on a number of heterogeneous, 
low-power, and loosely-coupled devices (i.e. for privacy constraints);
\item training code that can be redistributed, possibly in form of binaries 
(i.e. to train models at customer's premises without exposing sensitive Python 
code).\label{aim:3}
\end{enumerate*}
\md{I would rephrase the rest of the paragraph as a clear stating about research contributions. -- PARTIALLY DONE, more radical change needed?}
To practically implement this vision are required a number of advancements, this work represents a first step towards:
\begin{enumerate}
\item a better theoretical understanding of the different strategies of data 
parallelism in deep neural networks;\label{goal:1}
\item a consistent way to compare different deployments and 
strategies.\label{goal:2}
\end{enumerate}
Issues related to point \ref{goal:1} will be presented, 
addressing some of them and discussing how it is possible to push further the 
model training efficiency; moreover, this paper will propose a design for a 
programming framework that would address point \ref{goal:2}.

Sec. \ref{sec:background} presents a survey of parallel techniques for deep 
neural network training, the next section provides a further exploration of some 
theoretical highlights that can be exploited to improve training scalability. 
Sec. \ref{sec:framework} presents a design for an upcoming data parallel 
training framework and, finally, sec. \ref{sec:conclusion} provides an outlook 
of the potential impact of the presented results as well as the opportunities.

\section{Background}\label{sec:background}
Performance issues in deep neural networks (DNNs) have been extensively 
investigated from many point of views: in particular it is possible to clearly 
discriminate between the training stage and the inference 
stage. The latter is usually characterised by smaller computational workloads 
that are, however, highly constrained by time, memory, and power consumption due 
to the deployment on portable devices that need predictions almost in real-time. 
This paper is focused on the former stage of deep neural network training.
\md{the rest of the paragraph sounds a bit underselling (as if this work was
``subsumed'' by \cite{Ben-NunDemystifyingParallelDistributed2018}),
maybe better to resort it to the research contributions -- DONE}
A comprehensive survey of the state of the art for parallel DNN training been done by Ben-Nun and Hoefler \cite{Ben-NunDemystifyingParallelDistributed2018}, it is among the goals of this paper to review a subset of the relevant work, providing a more critical insight.

To further focus the research scope of this work, it is useful to highlight the 
main categorization of parallel training: namely \emph{data parallelism} vs. 
\emph{model parallelism}. Data parallelism focuses on distributing partitions of 
training data among workers, that cooperate to train replicas of the same model; 
model parallelism involves the partition of the model computation graph among 
different workers, that train different parts of the same model instance. While 
the latter (including layer pipelining) has been proved to be an efficient way 
to improve the performance of DNN training 
\cite{DeanLargeScaleDistributed2012,NgiamTiledconvolutionalneural2010,
ChenPipelinedBackPropagationContextDependent2012,
DengScalablestackinglearning2012}
it can be argued that its capacity to scale beyond the single machine is 
limited by the higher frequency of communications with respect to data parallelism, especially if the distributed workers are loosely coupled (i.e. cloud 
instances without dedicated interconnection, edge devices). Moreover, model 
parallelism can be used transparently within a distributed data parallel set-up 
to improve node-level performance, hence it represents an orthogonal direction 
of improvement with respect to data parallelism. In fact, this aspect is not 
explored in this work, but it can be quickly added to the data parallel 
strategies discussed later as a further optimization, without impacting the 
following discussion. \ma{you did not talked about data parallelism}

\subsection{Mathematical notation}\label{sec:math}
Despite the many attempts to implement different optimizations strategies 
\cite{SchmidhuberDeeplearningneural2015}, back-propagation 
\cite{LeCunDeeplearning2015,
WerbosApplicationsadvancesnonlinear1982,
LeCuntheoreticalframeworkbackpropagation1988,
LeCunEfficientBackProp1998} with some flavour of gradient descent 
\cite{BengioPracticalRecommendationsGradientBased2012} is still the most popular 
way to train deep neural networks, mostly due to its 
high efficiency on modern architectures like GPUs 
\cite{RainaLargescaleDeepUnsupervised2009}. This section present some useful 
notation for gradient descent-based neural network training.

For the rest of this section it will be considered that a dataset $X = \lbrace 
\mathbf{x_1}, \dots, \mathbf{x_n}\rbrace$, is used to train a neural network 
represented here as a collection of parameters (\emph{weights}) $\wvec = \lbrace 
w_1, \dots, w_m\rbrace$. Hereafter, neither the network type 
and topology (i.e. convolutional, recurrent, number of hidden layers) nor the 
input dimensionality and shape are considered relevant, as the formalism is generally applied to all of them. Mini-batch gradient descent 
\cite{OrrRemovingNoiseOnLine1997,
MollerSupervisedlearninglarge1992,
LeCunEfficientBackProp1998} has quickly became the standard, combining the 
faster convergence of Stochastic (\emph{on-line}) Gradient Descent (SGD) 
\cite{BottouTradeoffsLargeScale2008,
Wilsongeneralinefficiencybatch2003,
BottouLargeScaleOnline2004}, with the more efficient computation of \emph{batch} 
gradient descent.
\md{I would anticipate the definition of micro-batch here (now is just before
Eq. 3), otherwise $i$ is not defined -- DONE}
The optimization step for training can be expressed as the following weight 
update, computed with respect to a mini-batch ${X_{(i,n_b)} = \lbrace
\mathbf{x_i}, \dots, \mathbf{x_{i+n_b-1}}\rbrace}$:
\begin{equation}\label{eqn:gradient}
w_k(t + 1) = w_k(t) - \dfrac{\eta}{n_b} \sum_{j=i}^{i+n_b-1} \dfrac{\partial 
L\left(\wvec(t), \mathbf{x_j}\right)}{\partial w_k}
\end{equation}
where $t$ represents the current gradient descent iteration (\emph{step}), 
$\eta$ is the so-called \emph{learning rate} that defines the size of the step 
to be taken in the direction of the steepest descent, and $\nicefrac{\partial 
L\left(\wvec, \mathbf{x_j}\right)}{\partial w_k}$\md{maybe $w_1$ should be $w_k$? -- DONE} is the partial derivative of 
the loss function of the neural network with respect to the weight $w_k$, when 
calculated on the training sample $\mathbf{x_j}$. The partial derivative is 
averaged over all the samples belonging to a given subset (the \emph{mini-batch}) of 
the training dataset of size $n_b$. It is useful to recall the definition of all 
the versions of gradient descent by means of the value of $n_b$:
\begin{itemize}
\item $n_b = 1$, stochastic gradient descent
\item $1 < n_b \ll n$, mini-batch gradient descent
\item $n_b = n$, batch gradient descent
\end{itemize}
Note that batch averaging, as opposite of just summing, has a non-trivial 
impact on the convergence of the training 
\cite{BengioPracticalRecommendationsGradientBased2012}. It is also useful to 
define the gradient for all the weights of the model as following
\begin{equation}
\nabla L(\wvec,\mathbf{x_j}) = \left( \dfrac{\partial L\left(\wvec, 
\mathbf{x_j}\right)}{\partial w_1}, \dots, \dfrac{\partial L\left(\wvec, 
\mathbf{x_j}\right)}{\partial w_m} \right)
\end{equation}
this represents the direction of steepest slope of the loss surface calculated 
with respect to $\mathbf{x_j}$ in the parameter's space ($L: \mathbb{R}^m 
\rightarrow \mathbb{R}$); it is trivial to obtain the gradient and the step with 
respect to the whole mini-batch $X_{(i,n_b)}$ as
\begin{align}
\dfrac{1}{n_b} \sum_{j=i}^{i+n_b-1} \nabla L(\wvec,\mathbf{x_j}) 
\overset{\underset{\mathrm{def}}{}}{=} \Delta L(\wvec,X_{(i,n_b)}) \nonumber\\
\wvec(t+1) = \wvec(t) - \eta \Delta L(\wvec,X_{(i,n_b)})
\end{align}

Equation \ref{eqn:gradient} represents the simplest form of mini-batch gradient 
descent. Several algorithms have been developed to improve the convergence rate 
of DNN training, a good review of them can be found in literature 
\cite{GoodfellowDeepLearning2016,
Ruderoverviewgradientdescent2016}. The key points of these evolved algorithms 
are:
\begin{enumerate*}
\item variable learning rate, $\eta \rightarrow \eta (t)$;
\item accounting for previous gradient steps (e.g. \emph{momentum} 
\cite{Qianmomentumtermgradient1999});
\item defining a different learning rate for each weight $\eta(t) \rightarrow 
\eta(t,w_k)$ (e.g. ADAM \cite{KingmaAdamMethodStochastic2014}).
\end{enumerate*}
These points have an impact on parallel training implementation that will be 
discussed later.

\subsection{Training parallelism}\label{sec:parallelism}
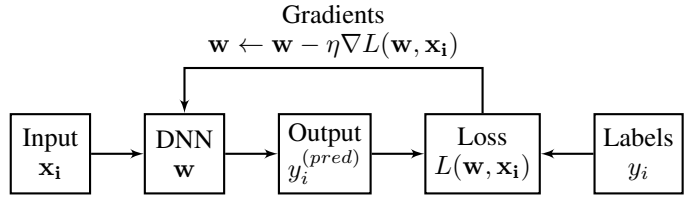
\begin{figure}[t]
\centering
\begin{tikzpicture}[auto,>=latex', thick]
  \node[draw,thick,minimum size=30pt,align=center] (inp) {Input\\$\mathbf{x_i}$};
  \node[draw,thick,minimum size=30pt, align=center, right =0.7cm of inp] (net) {DNN\\$\mathbf{w}$};
  \node[draw,thick,minimum size=30pt, align=center, right =0.7cm of net] (out) {Output\\$y_i^{(pred)}$};
  \node[draw,thick,minimum size=30pt, align=center, right =0.7cm of out] (loss) {Loss\\$L(\mathbf{w}, \mathbf{x_i})$};
    \node[draw,thick,minimum size=30pt, align=center, right =0.7cm of loss] (label) {Labels\\$y_i$};

  \draw [->] (inp.east) -- (net);
  \draw [->] (net.east) -- (out);
  \draw [->] (out.east) -- (loss);
  \draw [->] (label.west) -- (loss);
  \draw [->] (loss.north) -- +(0,0.6) -| node[above,pos=0.25,align=center] {Gradients\\$\mathbf{w}\leftarrow\mathbf{w}-\eta\nabla L(\mathbf{w}, \mathbf{x_i})$} (net.north);
\end{tikzpicture}
%
%
%
\caption{Back-propagation diagram for on-line gradient 
descent.}\label{fig:backprop}
\end{figure}

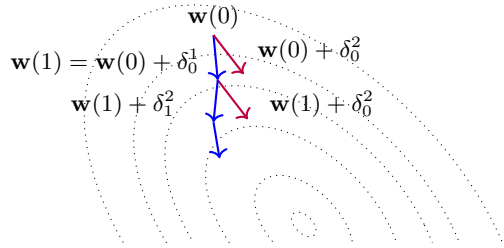
\begin{figure}[t]
\centering
\begin{tikzpicture}[scale=0.8,samples=100]
		\begin{scope}
			\clip(-5,-0.3) rectangle (4,4);
			\draw[dotted] plot[domain=0:360] ({cos(\x)*sqrt(20/(sin(2*\x)+2))},{sin(\x)*sqrt(20/(sin(2*\x)+2))});
			\draw[dotted] plot[domain=0:360] ({cos(\x)*sqrt(12/(sin(2*\x)+2))},{sin(\x)*sqrt(12/(sin(2*\x)+2))});
			\draw[dotted] plot[domain=0:360] ({cos(\x)*sqrt(8/(sin(2*\x)+2))},{sin(\x)*sqrt(8/(sin(2*\x)+2))});
			\draw[dotted] plot[domain=0:360] ({cos(\x)*sqrt(4/(sin(2*\x)+2))},{sin(\x)*sqrt(4/(sin(2*\x)+2))});
			\draw[dotted] plot[domain=0:360] ({cos(\x)*sqrt(1/(sin(2*\x)+2))},{sin(\x)*sqrt(1/(sin(2*\x)+2))});
			\draw[dotted] plot[domain=0:360] ({cos(\x)*sqrt(0.0625/(sin(2*\x)+2))},{sin(\x)*sqrt(0.0625/(sin(2*\x)+2))});
			
			\draw[->,blue,thick] (-1.5,3.15) to (-1.43,2.4);
			\draw[->,purple,thick] (-1.5,3.15) to (-1.0,2.5);
			\draw[->,purple,thick] (-1.43,2.4) to (-0.93,1.75);
			\draw[->,blue,thick] (-1.43,2.4) to (-1.5,1.7);
			\draw[->,blue,thick] (-1.5,1.7) to (-1.4,1.1);
			
			\node at (-1.5,3.45){\small $\wvec(0)$};
			\node at (-3.3,2.7){\small $\wvec(1) = \wvec(0)+\delta_0^1$};			
			\node at (0.1,2.9){\small $\wvec(0)+\delta_0^2$};
			\node at (0.3,2.0){\small $\wvec(1)+\delta_0^2$};
			\node at (-3.0,2.0){\small $\wvec(1)+\delta_1^2$};
		\end{scope}
\end{tikzpicture}
\caption{Gradient descent in $\wvec$ space. $\delta_i^j = - \eta\nabla 
L(\wvec(i),\mathbf{x_j})$ represents the gradient calculated on the weights 
updated up to step $i$, based on sample (or mini-batch) $\mathbf{x_j}$. 
Therefore, the red update based on $\delta_0^2$ is outdated with respect to 
$\wvec(1)$, but its impact is not necessarily detrimental to the training. The 
target function is $L: \mathbb{R}^m \rightarrow 
\mathbb{R}$.}\label{fig:grad_descent}
\end{figure}

When considering the whole \emph{feed-forward/back-propagation} 
\cite{LeCunDeeplearning2015} training process, it is important to remark that it 
is, to some extent, intrinsically sequential. Figure~\ref{fig:backprop} and 
equation~\eqref{eqn:gradient} show how the gradient value depend on the present 
$\wvec(t)$ configuration and how its application through back-propagation 
produces a new configuration $\wvec(t+1)$: the new weights represent a data 
dependency for the feed-forward step for sample $x_{i+1}$, that must come 
strictly after the back-propagation, otherwise the gradient would be calculated 
based on outdated (\emph{stale}) weights.
In principle this prevents any kind of input sample-based parallelism while, in 
fact, this is true strictly for on-line SGD: the concept itself of batch (or 
mini-batch) gradient descent involves parallelism. The gradients related to all 
the samples in the (mini-)batch are computed based on the same value of $\wvec$ 
and, possibly, at the same time. It is worth noting that the data dependency 
depicted in figure~\ref{fig:backprop}, is introduced by on-line training 
algorithm and not by the problem itself, hence there is room to relax this 
dependency, either with mini-batches or with more sophisticated techniques that 
relax the dependencies \emph{between} mini-batches. 
Figure~\ref{fig:grad_descent} exemplifies a possible behaviour of SGD on a loss 
surface: it is not necessarily true that using always the most recent gradient 
leads to the best training accuracy, even the red update could end up to good 
loss minimum. In this sense is important to remember that the loss surface of 
DNNs is highly non-linear and difficult to describe globally 
\cite{ChoromanskaLossSurfacesMultilayer2015,
KeskarLargeBatchTrainingDeep2016}: a certain amount of noise and randomness 
associated to the gradient descent can be beneficial to the training outcome in 
terms of generalization. The next subsections will describe how this behaviour 
can be exploited to introduce some degree of parallelism into the training 
process.

\subsubsection{Synchronous parallelism}\label{sec:intro_sync}

As stated before\md{Guy's golden rule: avoid back-pointing references until really necessary -- DONE},
mini-batch gradient descent combines the best 
of both on-line and batch training; in particular, the fact that it can be 
expressed as a chain of matrix-matrix multiplication (GEMMs) 
\cite{LeCunEfficientBackProp1998} that allow for a very efficient implementation 
on multicore CPUs and GPUs 
\cite{BengioPracticalRecommendationsGradientBased2012}, enabled a wide adoption 
of deep learning due to the better training feasibility.
From the parallel computing point of view, mini-batches represent the most 
elementary approach to what is called \emph{synchronous data parallel} training, 
as a global synchronization happens at the end of each mini-batch. 

The amount of available parallelism depends on the size of the mini-batches, 
that in turn affects the convergence of the training. Apart from avoiding the 
extreme cases of on-line and batch gradient descent, the choice of the right 
mini-batch size is not trivial, and there is interaction with other 
hyper-parameters, like the  learning rate, as widely discussed in literature 
\cite{KeskarLargeBatchTrainingDeep2016,BottouOptimizationMethodsLargeScale2016,
JastrzebskiThreeFactorsInfluencing2017,
SmithDonDecayLearning2017,
ChenRevisitingDistributedSynchronous2016,
MastersRevisitingSmallBatch2018} often concerning the \emph{linear scaling} of 
$\eta$. In principle, larger mini-batches allow to process more samples per unit 
of time, while the convergence can be hindered if the size is too large.

Mini-batch parallelism is usually exploited by means of parallel GEMMs on 
suitable architectures \cite{RainaLargescaleDeepUnsupervised2009,
BergstraTheanoDeepLearning2011,
ChetlurcuDNNEfficientPrimitives2014}. However, recent works 
\cite{GoyalAccurateLargeMinibatch2017,
ChoPowerAIDDL2017,
AkibaExtremelyLargeMinibatch2017} have demonstrated that it is possible to push the mini-batch 
size further than previously expected without affecting the model convergence. 
These works leverage distributed GPU architectures in order to allocate and 
efficiently compute such large mini-batches, while relying on an 
\emph{all-reduce} communication pattern to perform the global synchronization. 
Ignoring the communication bottlenecks that will be discussed in 
Sec.~\ref{sec:compression}, it can be argued that this approach is 
problem-specific and can not always be pushed as far as 
\cite{GoyalAccurateLargeMinibatch2017} suggests. In fact, smaller mini-batches 
($\sim32$) provide usually better generalization performance 
\cite{LeCunEfficientBackProp1998,
KeskarLargeBatchTrainingDeep2016,MastersRevisitingSmallBatch2018}. This induces 
a granularity problem: smaller batches can be effectively computed only if the 
size of the network and the complexity of the individual data sample (e.g. large 
RGB picture vs. small array of numerical data) are large enough to saturate the 
given platform even with only few samples being processed concurrently. This 
issue can heavily affect the capability of certain models to scale on large 
distributed clusters.
A further issue is the so-called \emph{batch normalization} (BN) 
\cite{IoffeBatchNormalizationAccelerating2015}, that introduces data 
dependencies between different samples among the same mini-batch, such that a 
full synchronization is required at each invocation of BN.

Parallelism at mini-batch level proved to be effective at node-level when 
implemented on GPUs, multi-core CPUs or other dedicate hardware (e.g. Google 
TPUs \cite{JouppiInDatacenterPerformanceAnalysis2017}); still, the scalability 
of its extension to distributed memory architectures is subject to a suitable 
problem granularity, that is far from being granted apart from specific 
problems.

Further parallel implementation of DNN training usually take mini-batch 
parallelism for granted, at least at node-level, considering mini-batches as 
atomic entities for which the data dependency defined in 
Figure~\ref{fig:backprop} exists. From this point of view, mini-batches can be 
considered the only truly synchronous kind of parallel training: while other 
strategies that will be presented in the next sections might involve 
synchronizations at certain stages, they necessarily relax the dependency 
between subsequent mini-batches. Indeed, in the rest of this paper mini-batches 
will be considered as atomic entities, that cannot be further divided. 
Synchronous \emph{distributed} parallelism at mini-batch level will also be addresses as 
\emph{large mini-batch} parallelism.

\subsubsection{Asynchronous parallelism}\label{sec:intro_async}
The success of momentum as a method to accelerate the training convergence, show 
that the information of previous gradients is definitely relevant even at the 
current iteration. Although the idea of trading gradient staleness for 
computational efficiency can be also related a posteriori to the usage of 
mini-batches, as highlighted by Masters and Luschi 
\cite{MastersRevisitingSmallBatch2018}, this notion has been at first exploited 
for what is defined \emph{asynchronous parallel training}. As the name suggests, 
this strategy involves multiple workers performing their own gradient descent 
for a certain amount of iterations, while their findings (i.e. new weights, 
accumulated gradients) are shared with other workers without a global 
synchronization at the mini-batch level.

There is a common categorization \cite{Ben-NunDemystifyingParallelDistributed2018}
between centralized and de-centralized implementations, as well as based the degree of 
model consistency achieved. The latter is a property of a given implementation 
that measures how different are the weights of each model replica at a certain 
instant of time, while the former categorization regards the usage of a 
centralized \emph{parameter server} to store a ``master copy'' of the model 
weights or, otherwise, to coordinate the exchange of gradients without a central 
authority. Sec.~\ref{sec:discussion} will further discuss these classifications.
Early notable implementations of asynchronous parallel gradient descent are
HOGWILD! \cite{NiuHOGWILDLockFreeApproach2011} and its deep learning-focused derivatives like Downpour SGD
\cite{DeanLargeScaleDistributed2012,ChilimbiProjectAdamBuilding2014}; followed 
by some other significant works \cite{PaineGPUAsynchronousStochastic2013,
StromScalableDistributedDNN2015,
ZhangStalenessawareAsyncSGDDistributed2015,
ZhengAsynchronousStochasticGradient2016a,
KeuperAsynchronousParallelStochastic2015,
HermansAccumulatedGradientNormalization2017,
LianCanDecentralizedAlgorithms2017}. Apart from the \emph{DistBelief} 
\cite{DeanLargeScaleDistributed2012} and Project Adam
\cite{ChilimbiProjectAdamBuilding2014} papers, that presented results previously 
not achievable and moved deep learning resolutely into the HPC domain, most of 
other works, while reporting solid scalability and timing results, were not able 
to provide a significant legacy. In fact, the dominating entries from DAWNBench 
\cite{ColemanDAWNBenchEndtoEndDeep2017}, at the time of writing, are still 
relatively small-scale, synchronous implementations.

While this review is far from being conclusive, it is possible to suggest some 
limitations that arguably prevented widespread adoption of asynchronous 
techniques. For instance, the added complexity of a parameter server or a 
sophisticated decentralized protocol might be perceived as not necessary since 
synchronous, all-reduce-based, parallelism has mostly satisfied the quest for deep learning 
scalability up to this point. Moreover, most of these works present asynchronous 
implementations of naive SGD, while the state of the art is moving to more 
sophisticated algorithms like ADAM \cite{KingmaAdamMethodStochastic2014}. Some 
effort in this directions exists \cite{HermansScalableDeepLearning2017}, as well 
as a prominent theoretical work \cite{MitliagkasAsynchronybegetsMomentum2016} 
that links gradient staleness to momentum; still, the literature is lacking a 
comprehensive analysis of the asynchronous behaviour of algorithm beyond SGD. 
Finally, results are usually reported as a collection of experiments on specific 
use cases, lacking a generalization effort that might help to understand the 
validity of the methodology. In this sense a relevant analysis has been 
performed by Lian et. al \cite{LianAsynchronousDecentralizedParallel2017}: the 
theoretical discussion of the convergence rate for an asynchronous, 
decentralized algorithm represent a good starting point for a performance 
analysis. However, it can be argued that the real life behaviour is affected by 
a large number of variables (e.g. weight update protocol,communication 
latencies, etc.) that 
prevent this model to fully describe the performance of a given implementation. These limitations, along with the lack of details on the code and framework used for experiments, lay the ground for a research that aims to fill the gap between sparse experimentation and mathematical modelling of convergence rates. 

\subsubsection{Other approaches}
Synchronous and asynchronous SGD are not the only ways to exploit concurrency in 
DNN training. \emph{Model averaging} 
\cite{PolyakAccelerationStochasticApproximation1992,
ZhangDeepLearningElastic2015,PoveyParalleltrainingDeep2014} allow concurrent 
model replicas to perform training independently up to a certain point (i.e. 
from several mini-batches to multiple epochs), then the weights are averaged 
among the different replicas. \emph{Ensemble learning} 
\cite{LeeWhyHeadsare2015,HintonDistillingKnowledgeNeural2015} performs the whole 
training on different model instances, then averages the predictions among them. 
As said before with respect to model parallelism, ensemble learning represents 
an orthogonal direction of improvement with respect to parallel gradient 
descent, hence it will not be discussed hereafter. On the other hand, model 
averaging is strictly related to the techniques presented in 
Sec.~\ref{sec:intro_sync} and \ref{sec:intro_async} and, while it is out of the 
scope of this paper to formally draw the connection, it will be investigated in 
the near future.

\subsubsection{Further parallelism issues}\label{sec:compression}
As said in Sec.~\ref{sec:intro_sync}, mini-batch parallelism tends to be 
performed within a single node, either in shared memory or distributed among 
multiple GPU. The computing horsepower provided by GPUs or other dedicated 
hardware is usually enough for most applications, still, there is the need to 
push the capability to train DNNs effectively beyond the single node. While 
large mini-batches and asynchronous techniques can be applied also within a 
single machine when the problem is small enough, representing an interesting 
research domain itself, they are born to be distributed; this raises a number of 
issues related to the communication of gradient updates.

The size of the gradient set ($\Delta L(\wvec,X_{(i,n_b)}$) for a state of the 
art DNN easily reaches a few hundred MB \cite{LinDeepGradientCompression2017}. 
This represents a serious bottleneck for distributed implementations and two 
main techniques are used to reduce the size of the gradient set to be 
transmitted: \emph{quantization} and \emph{sparsification}. The former intends 
to reduce the precision of the gradient representation in order to reduce its 
overall size and it is demonstrated that this technique works up to 1-bit 
representation \cite{Seide1BitStochasticGradient2014,
StromScalableDistributedDNN2015}; the latter exploits the sparsity that 
naturally occurs in DNN gradients, where most of the components are zero or 
almost zero. In this way the array gradient component can be represented as 
sparse and compressed with well-known techniques 
\cite{StromScalableDistributedDNN2015}. A more recent work 
\cite{LinDeepGradientCompression2017} also includes momentum in the discussion 
and presents interesting results. Also in this case, apart from the 1-bit 
quantization provided by Microsoft CNTK 
\cite{YuIntroductionComputationalNetworks2014}, the frameworks used are not 
mentioned nor the code is made available.

More methodologies can be exploited to enhance the performance of distributed 
training, like the optimization of the all-reduce pattern required by the large 
mini-batch training or the overlapping of computation and communication during 
training. Even if these techniques fall more in the domain of the implementation 
details than in the field of parallel training algorithms, they play a 
non-negligible role in the overall training performance: this paper highlights 
the need of a general purpose framework that provides the tools to experiment 
with existing techniques at different levels (i.e. asyncronous vs. synchronous, 
different communication patterns, quantization, etc.), as well as defining and 
testing new ones. Sec.~\ref{sec:framework} will discuss the requirements for 
such framework.

\section{Theoretical discussion}\label{sec:discussion}
Assuming that using very large mini-batches is not suitable for any application, 
end-to-end training performance can be improved at two distinct
levels:
\ma{non tanto chiari, forse e' necessariod are un titolo ai due livelli, ci sono 2
  sottolivelli per ogni livello?}
\begin{enumerate}
\item at node level\label{enum:theory1}
\begin{itemize}
\item by implementing tensor operations in back-propagation even more 
efficiently;
\item by developing new dedicated hardware that is better suited to 
handle small mini-batches;
\end{itemize}
\item at distributed level\label{enum:theory2}
\begin{itemize}
\item by improving parallel gradient descent without falling back-on large mini-batches;
\item by developing a different optimization strategy that exploits 
parallelism better than gradient descent.
\end{itemize}
\end{enumerate}
Point \ref{enum:theory1} is being researched actively 
\cite{VasilacheTensorComprehensionsFrameworkAgnostic2018,
MarkidisNVIDIATensorCore2018} and it is clearly out of the scope of this paper. 
Also the development of algorithms that departs completely from gradient descent 
is an interesting topic, still this work is focused on improving on parallel 
gradient descent. In this sense it is possible to show that, despite usually 
being treated as different approaches, all the techniques discussed in 
Sec.~\ref{sec:intro_sync} and \ref{sec:intro_async} can be placed on a spectrum 
of communication completeness, namely the property of parallel implementation to 
distribute each gradient update from each worker to all the other workers, 
regardless of the time at which this happens. It is indeed possible to argue 
that the model consistency spectrum usually proposed 
\cite{Ben-NunDemystifyingParallelDistributed2018}, provides limited insight to 
understand what happens to model replicas in implementations presented in 
previous works.
\begin{figure*}[t]
\centering
\begin{tikzpicture}[scale=0.9,font=\small,->,level 1/.style={sibling distance = 20em},
level 2/.style={sibling distance=10em},
level 3/.style={sibling distance=4em},
  every node/.style = {align=left}, edge from parent path={(\tikzparentnode.south) -- ++(0,-1em) -| (\tikzchildnode.north)}]
  \node [draw] {$\wvec_0$}
    child { node (a) {$\wvec_1^A = \wvec_0 + \delta_1^A(\wvec_0)$}
    	child { node {$\wvec_1^A + \delta_2^A(\wvec_1^A)$}
    	edge from parent node[above,font=\scriptsize] {Not received} }
    	child { node {$\wvec_1^A +\delta_1^B(\wvec_0)$\\$+\delta_2^A(\wvec_1^A+\delta_1^B(\wvec(0))$} 
    	edge from parent node[above,font=\scriptsize] {Received} }
    edge from parent node[above,font=\footnotesize] {Worker A} }
    child { node (b) {$\wvec_1^B = \wvec_0 + \delta_1^B(\wvec_0)$}
    	child { node {$\wvec_1^B + \delta_2^B(\wvec_1^B)$}
    	edge from parent node[above,font=\scriptsize] {Not received} }
    	child { node {$\wvec_1^B + \delta_1^A(\wvec_0)$\\$+\delta_2^B(\wvec_1^B+\delta_1^A(\wvec_0))$}
    	edge from parent node[above,font=\scriptsize] {Received} }
	edge from parent node[above,font=\footnotesize] { Worker B} };

\path (a) -- (b) node [midway,font=\large] {$\stackrel{\delta ?}{\myrightleftarrows}$};
\draw[->] (-6.8,0) to (-6.8,-3);
\node at (-6.4,0){time};
\end{tikzpicture}
\caption{Diagram of weights update between two workers. $\wvec_0$ is the common starting configuration. Assuming that all the updates that are not immediatly applied are queued somewhere, commutativity and associativity of vector sum guarantee that A and B will always be consistent once the queues are emptied.}\label{fig:consistency1}
\end{figure*}
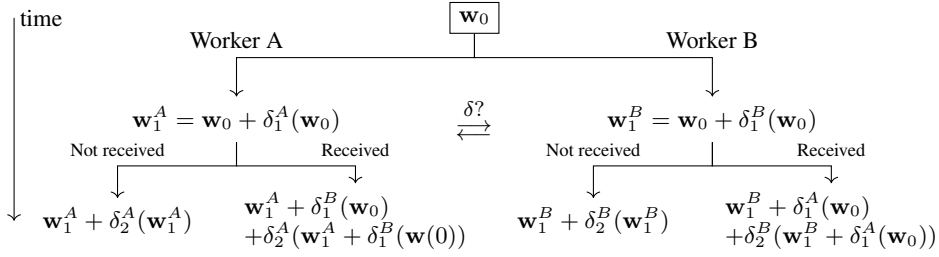
A statement can be formulated in this sense that, while being quite na\"ive, 
it is still important to understand the behaviour of model replicas
\begin{statement}\label{theo:consistency}
Assuming mini-batch SGD without momentum in a distributed setting, if 
all the gradient updates (communications) are delivered to all the workers, 
regardless of the delay, all the model replicas will be consistent.
\md{not clear why the following if clause is needed:
isn't true that guaranteeing all the communications already guarantees global synchronization? -- DONE right, I reformulated this a couple of times and I messed up}
\end{statement}
   Figure~\ref{fig:consistency1} presents 
the diagram of subsequent gradient updates for 2 workers: using commutativity 
and associativity of the vector sum that represent the gradient update, it is 
trivial to prove that, if an event triggers the application of all the pending updates (e.g. a global synchronization), whatever is the state of both workers before the event, their state will be consistent afterwards.
Of course statement~\ref{theo:consistency} does not hold 
if, for instance, updates not yet received are 
simply dropped, instead of accumulated. Moreover, it must be highlighted that 
having consistent model replicas does not mean that the result is the same as the 
sequential implementation, but only that all the model replica will agree on the 
value of $\wvec$ at a certain time. It is also important remark that consistency 
is not implied at any given moment, but it is always achieved as most of the strategies proposed either accumulate all the updates in 
a parameter server or require a synchronization at each epoch 
\cite{StromScalableDistributedDNN2015} or both.

In this sense there is also no need to distinguish between centralized and 
de-centralized set-ups if the communication is complete; in fact there it becomes 
only matter of implementation to choose the approach, while the model consistency 
is granted. While a centralized parameter server can simplify the measurement of 
gradient staleness, it is still possible to envision a distributed system that 
takes staleness into account.

This discussion is relevant as our goal is to exploit more parallelism without 
resorting to large mini-batch training; however, workers 
in figure~\ref{fig:consistency1} always go through the \emph{received} branch the outcome is, not surprisingly, exactly equal to the large mini-batch strategy. Less trivially, it is possible to 
figure that this is exactly what happens in an homogeneous, de-centralized set-up, where 
the load is perfectly balanced and updates are broadcast by each worker to all 
the others \cite{StromScalableDistributedDNN2015}, making an asynchronous solution not 
different from a synchronous one. Of course it can be argued that not enforcing 
explicit synchronization can benefit scalability on very large-scale deployment, 
however, it does not benefit the training as it is bound to an approximation of 
very large mini-batches.

It is useful at this point to define a new spectrum to discriminate between 
strategies:
\begin{enumerate}
\item Synchronous communication (large mini-batches)
\item Complete communication with bound delay (stale-synchronous 
\cite{ZhangStalenessawareAsyncSGDDistributed2015})
\item Complete communication with unbound delay (Downpour SGD 
\cite{DeanLargeScaleDistributed2012})
\item Partial communication (\cite{RamAsynchronousgossipalgorithms2009,
HoeflerCorrectedGossipAlgorithms2017})\label{point:partial}
\end{enumerate}
\md{``authors' opinions'' sounds a bit underselling, I think the assertion is quite well supported by the discussion above}
It is important to remark that, when applied in an homogeneous environment with high-bandwidth, low-latency interconnection (i.e. any common HPC set-up), the first three points are not 
significantly distinguishable in terms of training convergence.
It is true that a centralized set-up with a parameter server forces a degree of asynchrony since 
gradient updates are queued, still this is more a limitation of the centralized 
implementation that a property of this strategy, moreover the centralized approach introduces an obvious bottleneck. Point~\ref{point:partial} would 
be, instead, a significant departure from large mini-batches, and its benefit on 
the training convergence should be definitely investigated, while its 
scalability can be expected to be almost linear in terms of samples processed 
per unit of time, as it is for most of the asynchronous implementations. 
Moreover, this approach would significantly benefit in loosely-coupled 
heterogeneous environments (e.g. \emph{edge}), where the communication is costly 
and unreliable. However, while partial communication has been explored theoretically in generic optimization context \cite{RamAsynchronousgossipalgorithms2009}, no deep learning-related investigation has been conducted.

It is clear that allowing partial communication definitely gives up on model 
consistency, even in the long run. The impact of this on the training must be 
better understood, as well as the policy to determine which model to choose as 
representative when the training ends. This last issue is also strictly related 
to the possibility to terminate some workers at any given time without impacting 
the overall convergence: this matter has been already discussed 
\cite{DeanLargeScaleDistributed2012}, but only from the  point of view of fault 
tolerance of the training system, not in terms of training accuracy. Finally, it 
is necessary to investigate the impact of partial communication when more 
sophisticated optimization algorithms are used in place of na\"ive SGD. Momentum 
arises implicitly when introducing stale gradients 
\cite{MitliagkasAsynchronybegetsMomentum2016}, but there is no clear 
understanding of what happens in case of incomplete communication, as well as 
for more sophisticated algorithms with variable learning rates. It is reasonable to expect that the discussion made for the synchronous case by Goyal et al. \cite{GoyalAccurateLargeMinibatch2017} on momentum correction and aggregation of gradients subject to momentum can be extended for asynchronous set-ups with also implicit momentum and investigation is in progress in this sense.

To wrap up the discussion, asynchronous gradient descent with partial 
communication seems a promising alternative to more popular methodologies. The 
next section will discuss the requirements of a framework that can enable 
efficient experimentation on this topic.

\section{FAST C++ framework}\label{sec:framework}
\md{I would clarify the state of the library/prototype/proposal at the beginning of the section}
This library is currently\footnote{\today} 
under development and not yet publicly available.
In order to provide a truly general purpose tool, as well as to exploit the 
peculiarities of the different deep learning frameworks available, the proposed 
FAST (Flexible (A)synchronous Scalable Training) approach intends to decouple 
the intra-node execution of the training from the parallel coordination of 
workers; in fact it is reasonable that the user desires to keep using its 
framework of choice (e.g. Tensorflow, PyTorch, MxNet).

\begin{figure*}[t]
\centering
\begin{tikzpicture}[scale=0.7,auto,>=latex', thick,node distance=0mm,
comp/.style={
  draw,
  text width=2.5cm,
  align=center,
  text height=12pt,
  text depth=7pt,
  inner sep=0mm,
  outer sep=0mm},
]
  \node[comp] (tf) {Tensorflow};
  \node[comp, right=of tf] (pytorch) {PyTorch};
  \node[comp, right=of pytorch] (mxnet) {MxNet};
  \node[comp, right=of mxnet] (other) {...};
  \node[comp, above right=of tf.north west, text width=10cm] (fastlow) {FAST tensor moving interface};
  \node[comp, above right=of fastlow.north west, text width=7.5cm] (fasthigh) {FAST high-level strategies};
    \node[comp, above right=of fasthigh.north west, text width=10cm,draw=none] (business) {C++ or Python business logic};
    
    \draw (business.south west) -- (business.north west);
    \draw (business.north west) -- (business.north east);
    \draw (business.north east) -- (fastlow.north east);

\end{tikzpicture}
\caption{FAST logical stack.}\label{fig:fast}
\end{figure*}
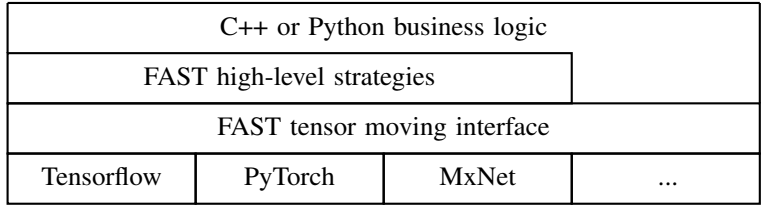

The main feature will be a general purpose tensor moving interface that allows 
the developer to send and receive any kind of tensor between model replicas, 
offering pre-defined compression and sparsification strategies. To achieve performance without loosing the flexibility and programmability required to a general purpose tool, a novel, state of the art, take on distributed shared 
memory will be leveraged \cite{17:gam:drocco:thesis}.
This interface 
will be both proposed to the user as-is to experiment novel approaches, and 
wrapped in a number of higher-level strategies based on literature, spanning the 
whole spectrum presented in Sec.~\ref{sec:discussion}. GPU-GPU communication for 
device-based tensors will be part of the implementation. Figure~\ref{fig:fast} presents the logical stack of components: this structure is also expected to 
allow better reproducibility of previous results while factoring out all the 
node-level performance optimizations, that are delegated to the underlying 
framework.

The library is designed from scratch with C++ training in mind, according to the 
aim of making training code redistributable, while potentially target training in 
production and keeping the overhead as low possible. However, due to the 
prevalence of Python for DNN training, Python wrappers will be provided 
compatible with selected frameworks.

\section{Conclusion and future work}\label{sec:conclusion}
It is very likely that the next breakthrough in training performance will either come from new dedicated silicon architectures or from theoretical advancements in optimizations techniques that departs from gradient descent. However, at this stage, the quest for training performance at scale has been met mostly by synchronous, large mini-batch, parallelism; unfortunately this strategy is heavily problem-dependent, moreover, it is not suitable for other platforms than conventional HPC clusters and tightly coupled cloud instances.

This paper endorses a departure from both synchronous and conventional asynchronous training, as they both perform similarly in terms convergence when working within a high-performance infrastructure. Instead, asynchronous training with sparse communication is expected to introduce a degree of randomization in the interleaving of updates coming from different mini-batches that represents a novelty with respect to large mini-batches and might arguably be beneficial to the training.

This approach would require an effort on both the theoretical and experimental side, in order to investigate the potential issues reported in Sec.~\ref{sec:discussion}. This work is currently taking place and tackles the issues related to model inconsistency that derives from partial communications, while the development of FAST library will allow to validate theoretical results on real models and datasets.

%

\bibliographystyle{IEEEtran}
\bibliography{deep_learning,distributed_systems,dnn_math,dnn_misc}

\end{document}